\documentclass[aps,prb,superscriptaddress,floatfix,twocolumn,amsmath]{revtex4-1}
\usepackage{bm,amsmath,amssymb,mathrsfs,dcolumn}
\usepackage{subfigure}
\usepackage{units}
\usepackage{hyperref}
\usepackage{mathtools}
\usepackage{wrapfig}
\usepackage{multirow}
\usepackage[usenames]{color}
\usepackage[export]{adjustbox}
\hypersetup{pdfborder=0 0 0,colorlinks=true,citecolor=blue,linkcolor=blue}
\newcommand{\angstrom}{\mbox{\normalfont\AA}}

\begin{document}

\title{Long-range order imposed by short-range interactions in methylammonium
lead iodide: Comparing point-dipole models to machine-learning force fields}

\date{\today}

\author{Jonathan Lahnsteiner}
\author{Ryosuke Jinnouchi}
\author{Menno Bokdam}
\email{menno.bokdam@univie.ac.at}
\affiliation{University of Vienna, Faculty of Physics and Center for Computational Materials Sciences, Sensengasse 8/12, 1090 Vienna, Austria}

\begin{abstract}
The crystal structure of the MAPbI$_3$ hybrid perovskite forms an intricate electrostatic puzzle with different ordering patterns of the MA molecules at elevated temperatures.  For this perovskite three published model Hamiltonians based on the point-dipole (pd) approximation combined with short-range effective interactions are compared to a recently developed machine-learning force field. A molecular order parameter is used to consistently compare the transformation of the anti-ferroelectric ordering in the orthorhombic phase upon raising the temperature. We show that the ground states and the order-disorder transition of the three models are completely different. Our analysis indicates that the long-range order in the low-temperature orthorhombic phase can be captured by pd-based models with a short cutoff radius, including the nearest and next-nearest neighbor molecules. By constructing effective atomic interactions the ordering can already be described within a $6$~\angstrom{} radius. By extracting the coupling energetics of the molecules from density functional theory calculations on MA$_x$Cs$_{1-x}$PbI$_3$ test systems, we show that the pd-approximation holds at least for static structures. To improve the accuracy of the pd-interaction an Ewald summation is applied combined with a distance dependent electronic screening function.
\end{abstract}

\maketitle

\section{Introduction}

\begin{figure}[!b]
 \centering
   \includegraphics[width=\columnwidth]{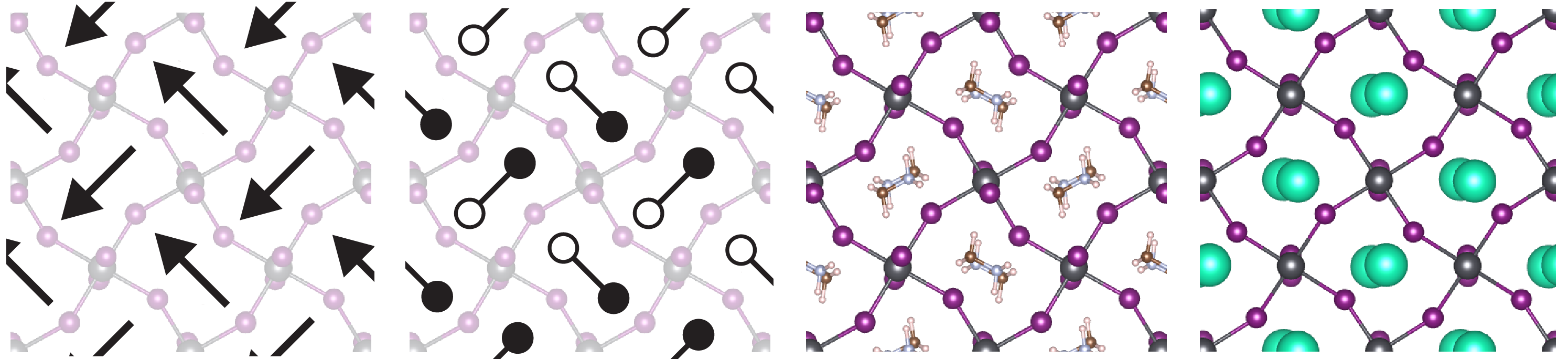}
 \caption{The low temperature orthorhombic structure of MAPbI$_3$ in the i) point-dipole model, ii) monopole model and iii) the machine-learning force field. Arrows indicate the point-dipoles $\mathbf{\hat{p}}$, and their monopole based equivalents are shown by the black \&{} white circles. The low temperature orthorhombic structure of CsPbI$_3$\cite{Sutton:acsel18} is shown in the last figure. }
 \label{structures}
\end{figure}

The finite temperature structure of the promising Methylammonium-Lead Iodide (MAPbI$_3$) hybrid perovskite solar cell material is characterized by a vast number of accessible macro-states related to the orientational degree of freedom of the Methylammonium (MA) molecule. Above $\sim333$~K, a cubic crystal structure is formed by PbI$_6$ octahedra enclosing MA molecules. Below $\sim150$~K, an orthorhombic lattice is formed, with molecules in a long-range ordered pattern. The MA$^+$Pb$^{2+}$I$_3^-$ system is held together by a mixture of covalent and ionic bonds, whereby the MA molecule is chemically decoupled from the PbI framework. Weak (hydrogen, van der Waals) bonds link the MA molecule to the framework\cite{Egger:jpcl14,Lee:cc15,Lee:sr16,Jingrui:prb16}, and the intrinsic dipole moment ($|\mathbf{p}|$) of the molecules couples them, even more weakly, to each other\cite{Frost:nanol14}. 
There is a large spread in the reported MA preferred orientations as well as in their rotation barriers (10-100~meV), see Ref.~\cite{Lahnsteiner:prm18} and references therein. On one thing all seem to agree, at room temperature and above the molecules show entropic disorder\cite{Poglitsch:jcp87} in their orientations. It is important to understand the ordering pattern (or lack thereof) of the MA molecules, since the orientation of the molecules are connected to various physical effects (screening, charge localization, polarons, Rashba) which affect electronic device properties\cite{Frost:nanol14,Ma:nanol14,Zheng:nanol15,Motta:natc15,Etienne:jpcl16,Neukirch:nanol16,Gong:jpcl16,Hu:jpcc17,Kang:jpcc17,Govinda:jpcl17}. Previous first-principles based molecular dynamics (FPMD) calculations have indicated that a soft long-range ordering pattern between neighboring MA molecules remains at room temperature \cite{Lahnsteiner:prb16,Lahnsteiner:prm18}. This order is a remnant of the highly ordered molecular pattern in the low-temperature orthorhombic phase. This order has been well resolved in experiment and is sketched in Figure~\ref{structures}. This pattern also shows the lowest energy in DFT calculations\cite{Menendez:prb14,Filip:natc14,Filip:prb14,Quarti:com14,Lee:cc15,Lahnsteiner:prb16}. However, the length and time scales required to model the finite temperature structure occurring in real devices exclude a straightforward application of FPMD methods. Therefore effective model Hamiltonians that describe the potential energy surface of MA molecules on a lattice could be of great help in understanding the local structure at  defects\cite{Simenas:jmcc18} and domain boundaries\cite{Ashhab:jap19}. The presence of the reorienting molecules makes it difficult to apply models used for other perovskites, like BaTiO$_3$\cite{Zhong:prb95}. Therefore point-dipole (pd) based model Hamiltonians\cite{Frost:aplm14,Leguy:natc15,Pecchia:nanol16,Motta:prb16,Simenas:jpcl17,Tan:acsel2017,Li:prb18,Jarvi:njp18,Ashhab:jap19}, a rigid rotor cluster expansion\cite{Thomas:prb18} and classical force fields\cite{Mattoni:jpcc15,Handley:pccp2017} have been designed. Often pd-models have been constructed in such a way that the heat capacity as function of temperature and/or phase transitions resemble experimental data. In some cases the coupling energies of certain inter-molecular ordering patterns have been fitted to DFT energies calculated for (relaxed) surrogates of the structure at different temperatures obtained from the experiment\cite{Weber:zfn78,Onoda-Yamamuro:jpcs90,Baikie:jmca:13,Stoumpos:ic13,Kawamura:jpsj02,Whitfield:sr16}. The local microscopic structure is in most experiments only observed on a time and space average. These surrogates show dynamic instabilities that manifest themselves as imaginary phonon frequencies in FP calculations\cite{Brivio:prb15,Leguy:pccp16}. This indicates that the higher temperature crystal phases are possibly entropically stabilized. When many different mirco-states of the molecular orientation are accessible, the related gain in entropy can shift the Gibbs free energy of the system such that the tetragonal and cubic phases are stabilized under certain temperature and pressure conditions.

In this work, we compare the inter-molecular structure at elevated temperature described by three recently published pd-based models\cite{Frost:aplm14,Leguy:natc15,Tan:acsel2017,Simenas:jpcl17} to a highly accurate machine learned force field (MLFF)\cite{Jinnouchi:prl19}. The MLFF has the advantage that it learns \textit{on-the-fly} during FPMD and generates ensembles that are practically indistinguishable from FPMD at much lower computational cost. By Monte Carlo calculations we study the change of this ordering pattern for the different models when the temperature is slowly raised. We show that the models have completely different ground states as well as excited states. Only one of the three considered models (Model III \cite{Simenas:jpcl17}) shows similar phase transitions as the MLFF. With DFT calculations on cubic MA$_x$Cs$_{1-x}$PbI$_3$ supercells we demonstrate that the pd-approximation of the MA molecule has a limited accuracy. By summing the dipole-dipole (d-d) electrostatic energy by means of an Ewald sum and by introducing a distance dependent model screening function we improve the accuracy. We show that the d-d coupling effectively splits the degeneracy of the states imposed by the short-range part of the Hamiltonian. These short-range interactions are the dominant energetic terms as can already be inferred from the experimentally observed long-range order in the low-temperature structure (Fig.~\ref{structures}). The molecules do not display a "striped" configuration (anti-ferroelectric aligned columns of head-to-tail dipoles) as would be expected for pure d-d interactions, but rather close to orthogonal orientations. Our analysis indicates that essence of this long-range order can be reduced to effective interactions within a $6$~\angstrom{} radius.

This paper is organized as follows. In Section~\ref{sec:model} we introduce the different model Hamiltonians, followed by a description of the applied Monte Carlo strategy and computational details in Section~\ref{sec:details}. The results are presented in Section~\ref{sec:results} and, for readability, split into several subsections. The paper is closed with a discussion in Section~\ref{sec:dis} and its conclusions are summarized in Section~\ref{sec:con}.

\section{Model Hamiltonians}
\label{sec:model}

Several model Hamiltonians describing the interactions of MA dipoles on a grid have been proposed in the last five years\cite{Frost:aplm14,Leguy:natc15,Pecchia:nanol16,Motta:prb16,Simenas:jpcl17,Tan:acsel2017,Li:prb18,Jarvi:njp18,Ashhab:jap19}. In these models an effective Hamiltonian splits the total interaction energy of all $N$ molecules in the cell in a short-range framework-dipole interaction and a long-range dipole-dipole interaction:  
$ H =  H_{\text{sr}} + H_{\text{lr}}$.
The interaction terms are often very similar in these models, therefore we will limit the analysis to three representatives. For the models here considered (hereafter referred to as, I\cite{Frost:aplm14,Leguy:natc15}, II\cite{Tan:acsel2017}, III\cite{Simenas:jpcl17}), the long-range part ($H_{\rm lr}$) is based on the electrostatic interaction energy of two point-dipoles
\begin{equation}
  U(\mathbf{p}_i,\mathbf{p}_j,\mathbf{ n }_{ij})=\frac{|\mathbf{p}|^2}{4\pi \varepsilon_{0}\varepsilon_{r}}\frac{1}{r_{ij}^{3}}\left(  \hat{\mathbf{p}}_{i}\hat{\mathbf{p}}_{j}  -3
                        (\hat{\mathbf{p}}_{i}\cdot\hat{\mathbf{ n } }_{ij} )(\hat{\mathbf{p}}_{j}\cdot\hat{\mathbf{ n } }_{ij} ) \right  ),
\label{dipdip}
\end{equation}
where $\hat{\mathbf{p}}_{i}$ and $\hat{\mathbf{p}}_{j}$ denote unit dipole vectors at different lattice sites. The distance between the dipoles is given by $r_{ij}=|\mathbf{n}_{ij}|$, and $\hat{\mathbf{n}}_{ij}=\mathbf{n}_{ij}/|\mathbf{n}_{ij}|$ is the unit displacement vector. The total long-range energy is then calculated by a sum over all neighbors within the interaction radius ($r_c$) and then summed over all lattice sites,
\begin{equation}
  H_{\text{lr}}=\frac{1}{2}\sum_{i=1}^{N}\sum_{j\in r_c}U(\mathbf{p}_i,\mathbf{p}_j,\mathbf{ n }_{ij}).
\label{long_range}
\end{equation}
All dipoles have the same dipole moment $|\mathbf{p}|$, but the relative permittivity $\varepsilon_{r}$ and the long-range cutoff ($r_{c}$) depend on the model. In model II the Ewald summation technique is used to compute $H_{\text{lr}}$. In the three models, the prefactor in Eq.~(\ref{dipdip}) is treated as an effective coupling constant for which an optimally tuned value is used. These values can be compared by $\varepsilon_{r}$ when we define the dipole moment of the MA molecule to $2.29$~D\cite{Frost:aplm14}. This results in $\varepsilon^{\rm I}_{r}=1$, $\varepsilon^{\rm II}_{r}=0.73$, $\varepsilon^{\rm III}_{r}=1.5$. \newline

\begin{figure}[!t]
 \centering
   \includegraphics[width=\columnwidth]{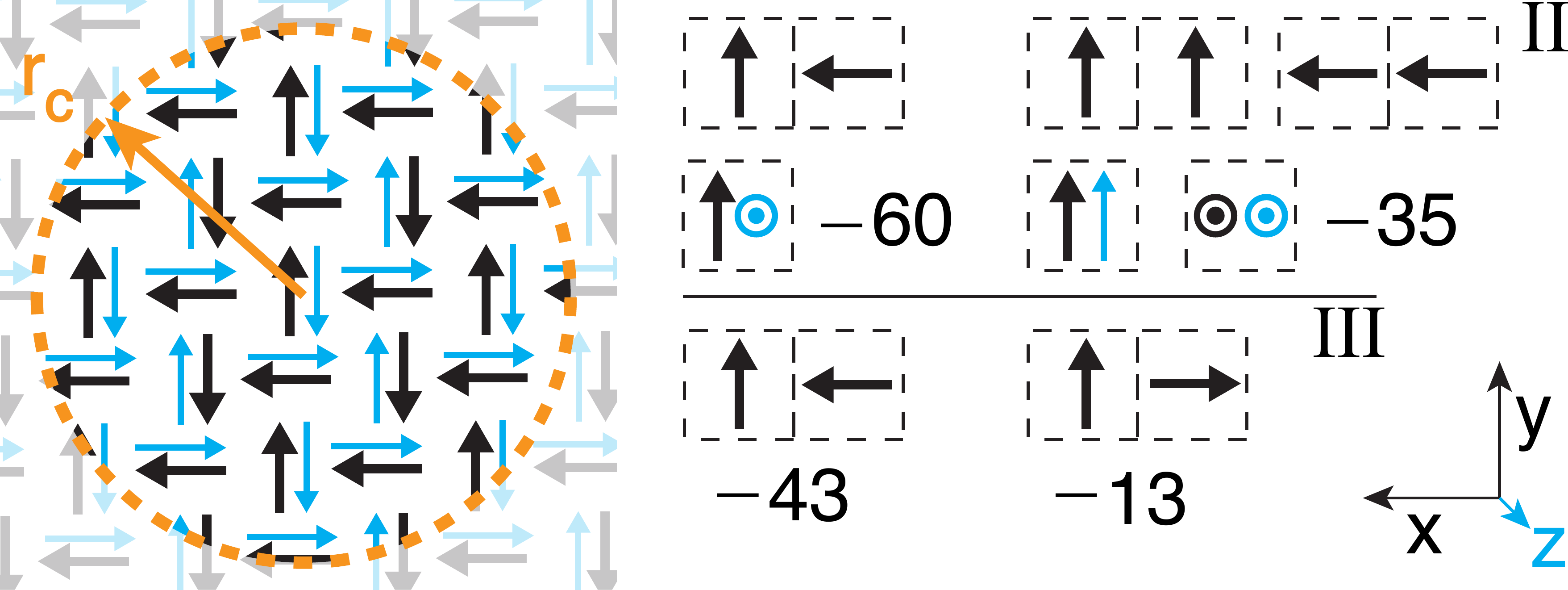}
 \caption{Point-dipole model on a cubic lattice with interaction radius ($r_c$). The pair-modes assigned in models II and III together with the corresponding penalty potentials $g \left( \hat{\mathbf{p}}_{i} , \hat{\mathbf{p}}_{j} \right )$ (in meV) are sketched on the right. The lower lying layer, into the paper z-direction, is colored in blue.}
 \label{shortr-fig}
\end{figure}

The short-range interaction ($H_{\rm sr}$) is differently modeled in models I, II and III. In model~I, it is described by a linear pair potential, 
\begin{equation}
 H^{\rm I}_{\text{sr}}=-K\frac{1}{2}\sum_{i=1}^{N}\sum_{j\in{\rm NN}}\hat{\mathbf{p}}_{i}\cdot\hat{\mathbf{p}}_{j},
 \label{local_leguy}
\end{equation}
\noindent
where the first sum runs over all lattice sites and the second sum over the nearest neighbors~(NN). The parameter $K=25$~meV approximates the strain on the framework and favors the parallel arrangement of dipoles. Both models II and III use a penalty potential of the form,
\begin{equation}
 H^{\rm II/III}_{\text{sr}} =\frac{1}{2}\sum_{i=1}^{N}\sum_{j\in{\rm NN}} g \left( \hat{\mathbf{p}}_{i} , \hat{\mathbf{p}}_{j} \right ).
 \label{tan_local}
\end{equation}
The strength of the penalty depends on the relative orientations of NN molecules as sketched in Figure~\ref{shortr-fig}. In model II, two interaction modes are described: the pair of molecules lies "flat" in a common plane $[(\hat{\mathbf{p}}_{i}\times{}\hat{\mathbf{p}}_{j})\cdot\hat{\mathbf{ n } }_{ij}=0]$ and (i) they are orthogonal and one of them points to the neighbor in the pair, then $g(\uparrow,\leftarrow)=-60$~meV, or (ii) they are pointing in the same direction, then $g(\uparrow,\uparrow)=-35$~meV.  For all other NN~configurations that can occur, $g=0$~meV. Model III, also describes only two pair modes, but these modes are more restricted: the pair of molecules lies orthogonal to each other in the $xy$-plane and one of them points (i) to the neighbor in the pair, then $g(\uparrow,\leftarrow)=-43$~meV, or (ii) away from the neighbor in the pair, then $g(\uparrow,\rightarrow)=-13$~meV. As before, for all other nearest neighbor configurations that can occur, $g=0$~meV. We note that the explicit preference for the $xy$-plane breaks the cubic symmetry of the Hamiltonian. In reality upon cooling from cubic to the orthorhombic phase, the molecules can condense in any of the cubic planes.

In regard to the short-range interaction of Eq.~(\ref{tan_local}) we would like to mention the thorough work of Li~\textit{et.al.}, who have systematically analyzed the "pair-mode" concept based on $4\times4\times4$ supercell DFT calculations\cite{Li:prb18}. The pair-modes used in models II and III also appear in their study as low energy configurations.

\section{Computational details}
\label{sec:details}

\textit{Monte-Carlo method.} All the model Hamiltonians are evaluated by Monte-Carlo methods based on the Metropolis-Rosenbluth algorithm \cite{Metropolis:jcp53}. $16\times16\times16$ dipole grids based on a cubic lattice ($a=6.3$~\angstrom) using periodic boundary conditions were simulated. The starting configuration is a minimal representation of the molecular order in the experimental low-temperature (orthorhombic) structure. It solely consists out of $x$ and $y$ axes orientated molecules as sketched in Fig.~\ref{structures}, and the next plane in $z$ direction has the same orientations anti-ferroelectrically aligned. Starting from the same configuration we calculate 20 Markov chains in parallel. The random number generator is initialized differently for every core to obtain different trajectories. Every chain equilibrates the system at the desired temperature by calculating $4\times10^6$ MC steps and then produces 25 equidistant samples spaced by $10\cdot16^3$ MC steps. Hereafter the temperature is increased by 10~K and the system is equilibrated again. This procedure is 60 times repeated, thereby raising the temperature from 0 to 600 K. Statistical averages are calculated over the chains. This setup has been carefully tested and shows converged order parameters.

Model I maps the orientational freedom of the dipoles on 26 discrete cubic orientations: $6\times[1,0,0]$ (axes), $12\times[1,1,0]$ (face diagonal) and $8\times[1,1,1]$ (room diagonal). To avoid a bias for an orientation, we have weighted the chances in the Monte Carlo steps accordingly. Models II and III are more restrictive and allow only for the six cubic axes directions. The cutoff-radii for the long-range interaction were set as in the respective publications, ie. $r^{\rm I}_c=3a$, $r^{\rm III}_c=2a$. For model II the Ewald technique as presented in Ref.~\onlinecite{FRENKELBOOK} is used instead of the real space summation.
\newline

\textit{Density Functional Theory.} For the first-principles calculations we use a plane-wave basis and the projector augmented wave (PAW) method\cite{Blochl:prb94b} as implemented in the {\sc vasp} 
code\cite{Kresse:prb93,Kresse:prb96,Kresse:prb99}. For the DFT calculations on the MA$_x$Cs$_{1-x}$PbI$_3$ test systems, the PBE (Perdew-Burke-Ernzerhof)\cite{Perdew:prl96} functional and standard {\sc vasp} pseudo-potentials are used with a 400~eV energy cut-off for the plane-wave basis. Gaussian smearing with $\sigma=0.05$~eV is used to broaden the one-electron levels. The Brillouin 
zone is sampled by a $2\times2\times2$ $\Gamma$-centered Monkhorst-Pack grid. The Kohn-Sham orbitals are updated in the self-consistency cycle until an energy convergence of $10^{-6}$~eV is 
obtained. $2\times{}2\times{}2$ CsPbI$_3$ super cells were constructed out 
of cubic unit cells with lattice constant $a=6.3$~\angstrom. On two sites, Cs atoms are replaced by Methylammonium (CH$_3$NH$_3$) molecules. All positions are kept fixed at their high symmetry cubic sites. The central molecule is rigidly rotated in spherical coordinates around the center of the C-N bond. No ionic relaxation was performed in these test systems.  \newline

\textit{Molecular Dynamics.} The machine-learned force field (MLFF) of MAPbI$_3$ as presented in Ref.~\cite{Jinnouchi:prl19} is used to generate MD trajectories. The MLFF was trained \textit{on-the-fly} during FPMD calculations with the SCAN (Strongly Constrained Appropriately Normed)~\cite{Sun:prl15} exchange-correlation functional, as its relative total energies agree well with many-body perturbation theory calculations in random phase approximation\cite{Bokdam:prl17}. The MLFF approach has the advantage that it speeds up the MD by three orders of magnitude, while retaining near FP accuracy. Note that the finite temperature structure of MAPbI$_3$ obtained with the MLFF is in excellent agreement with traditional FPMD~\cite{Lahnsteiner:prb16,Lahnsteiner:prm18}. Further computational details are presented in Ref.~\cite{Jinnouchi:prl19}. Here we have calculated  a 600~ps long trajectory for a $4\times4\times4$ supercell (768 atoms) in the $NPT$ ensemble. The temperature was linearly increased from 100 to 400~K using a Langevin thermostat\cite{Allen:book91}. Since the heating trajectory shows signs of hysteresis, we have calculated the averaged molecular order parameter on constant temperature MDs. In these calculations snapshots of the heating run were equilibrated for 0.2-1.2~ns, depending on the crystal phase and temperature.


\begin{figure*}[!t]
\centering
\includegraphics[width=2.07\columnwidth]{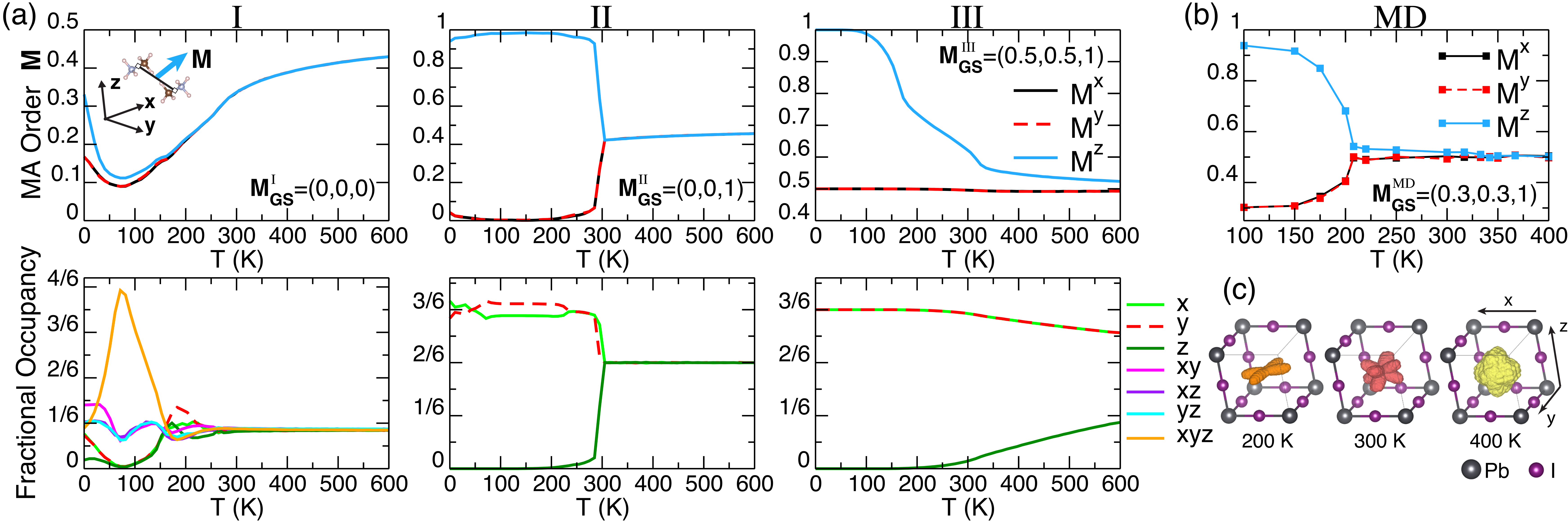}
\caption{(a) Molecular order parameter ($\mathbf{M}$) and fractional occupancy of orientational states as function of temperature for models I-III. Systems (I-III) are initialized at 0~K in state $\mathbf{M}_{\rm init}=(0.5,0.5,1)$ and the true ground state ordering is indicated by $\mathbf{M}_{GS}$. (b) The squares show the average order $\mathbf{M}$ obtained in constant temperature molecular dynamics (MD). (c) Three-dimensional polar distribution of the molecular C-N axes at three different temperature regimes in the MD. }
\label{molorder}
\end{figure*}

\section{Results}
\label{sec:results}
The results of this study are presented in four subsections. We start by introducing an order parameter for the molecules that is used to compare models I-III with MD. Hereafter we study the working of the short-range and long-range interactions separately.

\subsection{Molecular Order Parameter}
In order to straightforwardly compare the ordering patterns of the MA molecules we calculate the molecular order parameter as briefly presented Ref.~\cite{Jinnouchi:prl19}. We assume an integer cubic grid $\mathbf{q}=(i,j,k)$ of dimensions $N=N_{x}\times N_{y}\times N_{z}$ that has on each site a unit vector $\mathbf{\hat{p}}_{\mathbf{q}}$ describing the orientation of the MA molecule. The order parameter, $\mathbf{M}=\left({\rm M}^x,{\rm M}^y,{\rm M}^z\right)$, is then defined as
\begin{equation}
 {\rm M}^\alpha = \frac{1}{N\pi} \sum_{\mathbf{q}}^{N}\arccos{(\mathbf{\hat{p}}_{\mathbf{q}}\cdot\mathbf{\hat{p}}_{\mathbf{q}+\mathbf{n}^\alpha})},
\end{equation}
where $\mathbf{n}^\alpha$ is the displacement vector to the neighboring site in the Cartesian direction $\alpha$. Typical order parameters in the three different crystal phases are: 
\begin{equation}
\begin{aligned}
&\mathbf{M}_{\rm ortho}=(0.3,0.3,1)\\
&\mathbf{M}_{\rm tetra}\,=(0.5,0.5,0.53)\\
&\mathbf{M}_{\rm cub}\,\,\;=(0.5,0.5,0.5).
\end{aligned}
\end{equation}
These values agree well with the experimental structures as long as the temperature is not too close to the phase transition temperature\cite{Jinnouchi:prl19}.

\subsection{Molecular ordering at finite temperature}
To compare the ordering patterns of the dipoles imposed by the model Hamiltonians, we have calculated large finite temperature ensembles by slowly heating up the experimental low temperature molecular configuration  by Monte Carlo simulations. In { the first row of Figure~\ref{molorder}(a) the change of the order parameter ($\mathbf{M}$) during heating is shown. The initial structure for the models is described by $\mathbf{M}_{\rm init}=(0.5,0.5,1)$. $\mathbf{M}_{\rm init}$ is the closest equivalent of $\mathbf{M}_{\rm ortho}$ described with only axes orientations of the molecules. Models I and II already break down this structure at 0~K by exothermic steps, indicating it is not the true ground state (GS). Model II switches directly to its GS, which is an anti-ferroelectric (AFE) striped pattern, $\mathbf{M}_{GS}^{\rm II}=(0,0,1)$. Model I exhibits a kinetic barrier from $\mathbf{M}_{\rm init}\rightarrow\mathbf{M}_{GS}^{\rm I}=(0,0,0)$ and reaches it only partially at 75~K. The cubic symmetry breaking (indicated by different components of $\mathbf{M}$) observed at low temperature is an artefact. Its GS is the fully polarized state with molecules ordered along one of the \textit{room diagonal} (xyz) orientations. Model III retains the initial ordering at low temperature and shows a gradual transition involving two steps to a cubic-like phase. The different behavior of model I compared to II and III does not come as a surprise, its Hamiltonian has cubic symmetry and has no short-range interaction that can stabilize "flat" molecular orientations in the $xy$, $xz$ or $yz$ planes. Models II and III do have such interactions, imposed by favoring certain pair configurations. In model II, these pairs are counted for independently in which plane they lie, therefore the observed symmetry breaking is imposed by the initial structure only.  

The second row of Fig.~\ref{molorder}(a) shows the fractional occupancy of each quantized molecular orientation. Model I distinguishes 26 orientations, model II and III only 6 orientations. These different degrees of freedom cannot be compared directly, but do give an intuitive picture of the structure as function of temperature. At high temperature all orientations become equally occupied. Note here that there are eight \textit{room diagonal} orientations (xyz), compared to four \textit{face diagonals} (xy) and two \textit{axes} (x) orientations, therefore these directions are weighted accordingly. To build a model that describes all three crystal phases by their typical molecular orientations, one minimally needs the axes basisset. The expected occupancy of the orientations would then be for orthorhombic: ($\nicefrac{1}{2},\nicefrac{1}{2},0$), tetragonal ($\nicefrac{1-\delta}{2},\nicefrac{1-\delta}{2},\delta$) and cubic: ($\nicefrac{1}{3},\nicefrac{1}{3},\nicefrac{1}{3}$).

\begin{figure}[!t]
\centering
\includegraphics[width=\columnwidth]{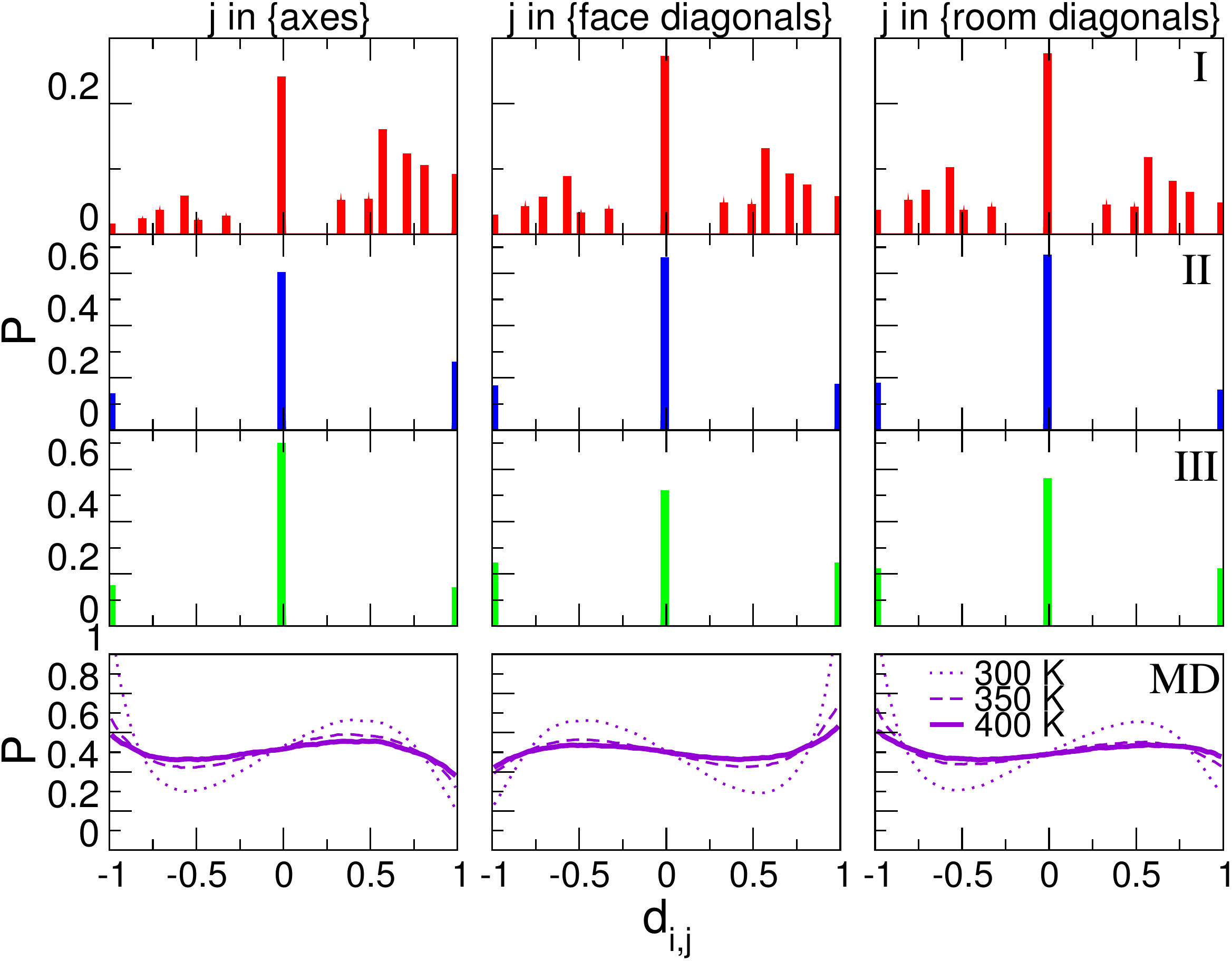}
\caption{Probability distributions of the relative molecular orientations ($d_{i,j}=\hat{\mathbf{p}}_i\cdot\hat{\mathbf{p}}_j$) in the first (axes), second (face diagonal) and third (room diagonal) nearest neighbor shells at 400~K. The MD distributions at 300 and 350~K are shown additionally.}
\label{hist}
\end{figure}

Using the order parameter as a common reference framework, we compare the models (Fig.~\ref{molorder}~a) to the MD results (Fig.~\ref{molorder}~b) in order to find the best working model. Both models I and II show different ordering compared to the MD already at 0~K. Their GS ordering patterns ($\mathbf{M}_{GS}^{\rm I}$,$\mathbf{M}_{GS}^{\rm II}$) are totally different compared to $\mathbf{M}_{\rm ortho}$. Model I does not mimic an orthorhombic/tetragonal to cubic phase transition, but shows a gradual changing ordering pattern with cubic symmetry. In model II, the barrier to leave the GS is roughly 300~K. The resulting transformation is manifested as a first-order phase transition (evidenced by a discontinuity in the first derivative of the order parameter) to the cubic phase. Model III shows qualitatively the best comparison with the MD. While heating up, the system undergoes two transitions at approximately 150 and 300~K. This is in agreement with the phase transitions obtained from peaks in the calculated heat capacity\cite{Simenas:jpcl17}. Similarly, the  MD shows two transitions at 200 and 350~K.  Each transition decreases the molecular order and brings it closer to $\mathbf{M}_{\rm cub}$. Furthermore, it has the correct GS ($\mathbf{M}_{\rm init}$). Still important differences between model III and the MD remain. Within the applied temperature window $\mathbf{M}_{\rm cub}$ is not reached, and both transitions are of second order. This is at odds with experimental data indicating a first order phase transition followed by a close to tri-critical transition\cite{Whitfield:sr16}. Interestingly, when turning off the long-range d-d interactions, model III breaks $\mathbf{M}_{\rm init}$ down to $\mathbf{M}_{\rm cub}$ at low temperature, even though the pair configurations only favor the $xy$-plane and can perfectly form the initial molecular order. We come back to this stabilizing effect in Section~\ref{subsC}.

Model III can capture the essence of the long-range order in the low temperature orthorhombic phase, although it fails to reproduce the more detailed molecular orientation. These orientations are illustrated in Figure~\ref{molorder}(c) by the three-dimensional distributions of the molecular C-N axis calculated from MD.  An iso-surface shows the outer shell of this distribution. The three different temperatures show the typical orientations of the molecules in the three phases. These missing orientations are also important to describe the probability distributions of the relative molecular orientation ($d_{i,j}=\hat{\mathbf{p}}_i\cdot\hat{\mathbf{p}}_j$). In Figure~\ref{hist} we see that all models show more structure at 400~K compared to the almost flat MD distributions. Furthermore, the MLFF captures the soft ordering that remains at room temperature and above, consistent with previous FPMD calculations\cite{Lahnsteiner:prb16}. In the first NN shell there is a broad peak at $d_{i,j}=\sim0.5$ and in the second at $d_{i,j}=\sim-0.5$ and so on. When using the axes basisset, both relative orientations are mapped to $d_{i,j}=0$.  Expanding the basis with the face- and room-diagonal orientations would allow us to reach a qualitative description for these relative orientations. To get realistic microscopic structures one would have to include the dominant orientations as they occur in the MD. 


\subsection{Splitting of degenerate ground states by dipole-dipole interaction}
\label{subsC}
In oder to clarify the origin of the long-range order, we analyze the short- and long-range interactions on the molecular orientations. As shown in Section~\ref{sec:model}, the short-range interaction is the energetically dominant term in all model Hamiltonians I, II and III. It is build up of NN pair interactions, and its ground state (GS) can be studied on a $2\times2\times2$ cubic grid with periodic boundaries. Using only the 6 \textit{axes} orientations we have calculated the total energies of all $6^8$ possible ordering patters. To illustrate the spread of these levels we have plotted the density of states (DOS) in Figure~\ref{dos}. We can roughly assess the influence of including the dipole-dipole coupling by calculating the NN interactions. In models II and III this increases the energy difference between the ground- and highest energy state ($\Delta{}E$), whereas it is unaffected in model I. More importantly, many of the degenerate states are split, thereby raising the total number of different energy states ($N_s$). This also holds for the states occupied at room temperature (roughly those between 0-25 meV). In this small system with only NN interactions we can determine the GS exactly; model I shows a fully polarized state with multiplicity $(\Omega)$ of 6, and model II shows an AFE striped pattern also with $\Omega=6$. Both GSs are different compared to the low temperature experimental ordering pattern of Fig.~\ref{structures}. We have summarized the characteristics of the models in Table~\ref{table1}, and we will now discuss model III.

\begin{figure}[!t]
\centering
\includegraphics[width=\columnwidth]{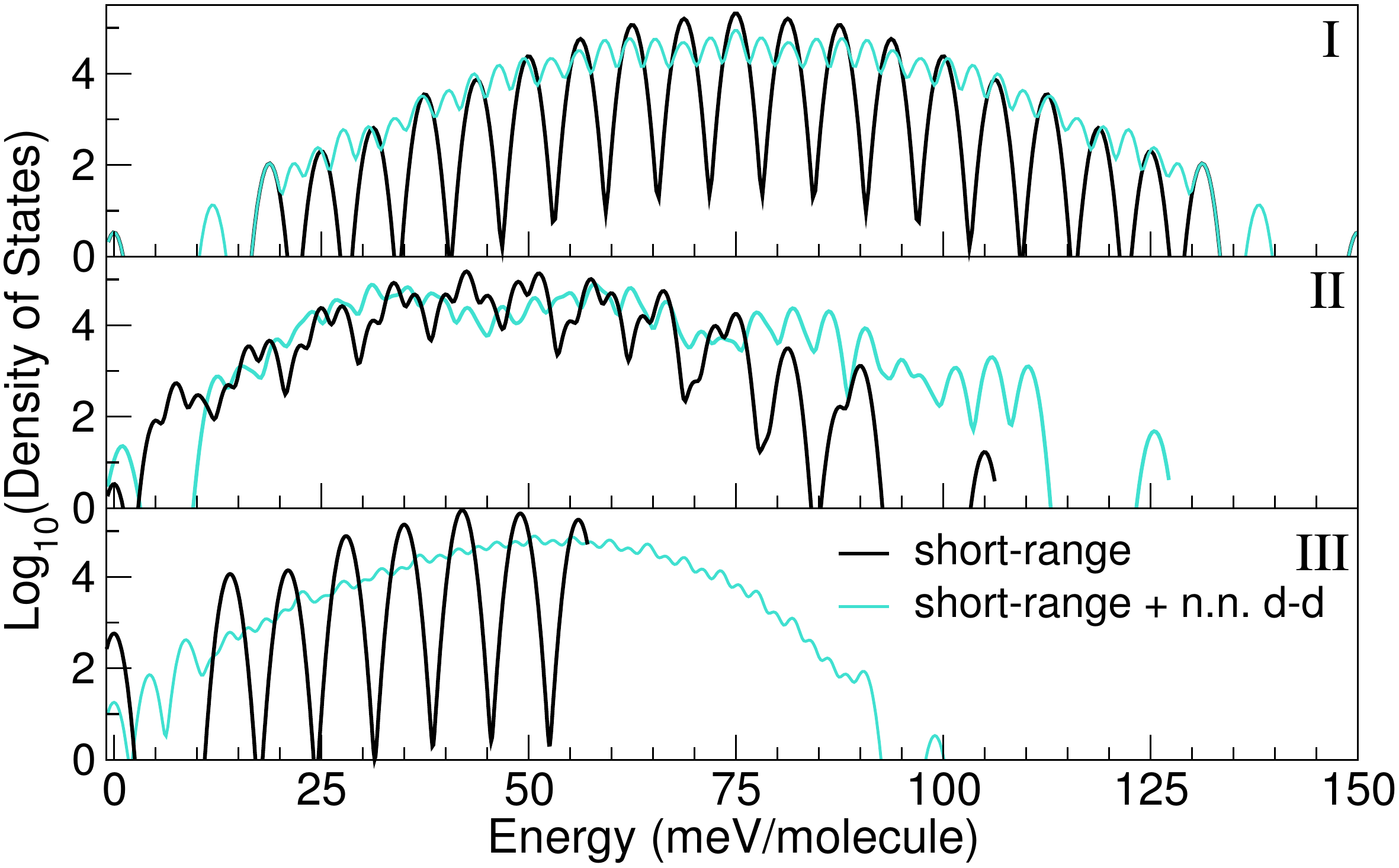}
\caption{Density of states (DOS) of models I-III on a $2\times2\times2$ cubic grid with only \textit{axes} orientations including periodic boundaries with (turquoise) and without (black) NN dipole-dipole coupling. The ground state energy is set to zero. Gaussian smearing ($\sigma=0.5$~meV) has been applied to broaden the $\delta$-functions at each energy level. Note that the DOS is normalized ($\int{}DOS(E)dE=6^8$), and its logarithm has been plotted. }
\label{dos}
\end{figure}

\begin{table}[!b]
\caption{Bandwidth of the models (meV/molecule) as expressed by the standard deviation ($\sigma$) of the DOS and the energy difference between the ground- and highest state ($\Delta{}E$). The multiplicity of the ground state ($\Omega$) and the total number of different energy states ($N_s$) are shown in the last columns.}
\label{table1}
\begin{ruledtabular}
\begin{tabular}{lcccc}
Model & $\sigma$ & $\Delta{}E$&  $\Omega$ & $N_s$ \\
\hline
Short-range I& 12 &150& 6 &21 \\
Short-range I + NN dipole-dipole& 15& 150 & 6 &95 \\
Short-range II& 12& 105 & 6 & 41\\
Short-range II + NN dipole-dipole& 17& 126 & 6 & 257\\
Short-range III& 9.3& 56 & 1024 & 8\\
Short-range III + NN dipole-dipole& 11& 99 & 32& 138\\
Short-range III + full dipole-dipole& -& - & 8& -\\
\end{tabular}
\end{ruledtabular}
\end{table}

\begin{figure}
\centering
\includegraphics[width=\columnwidth]{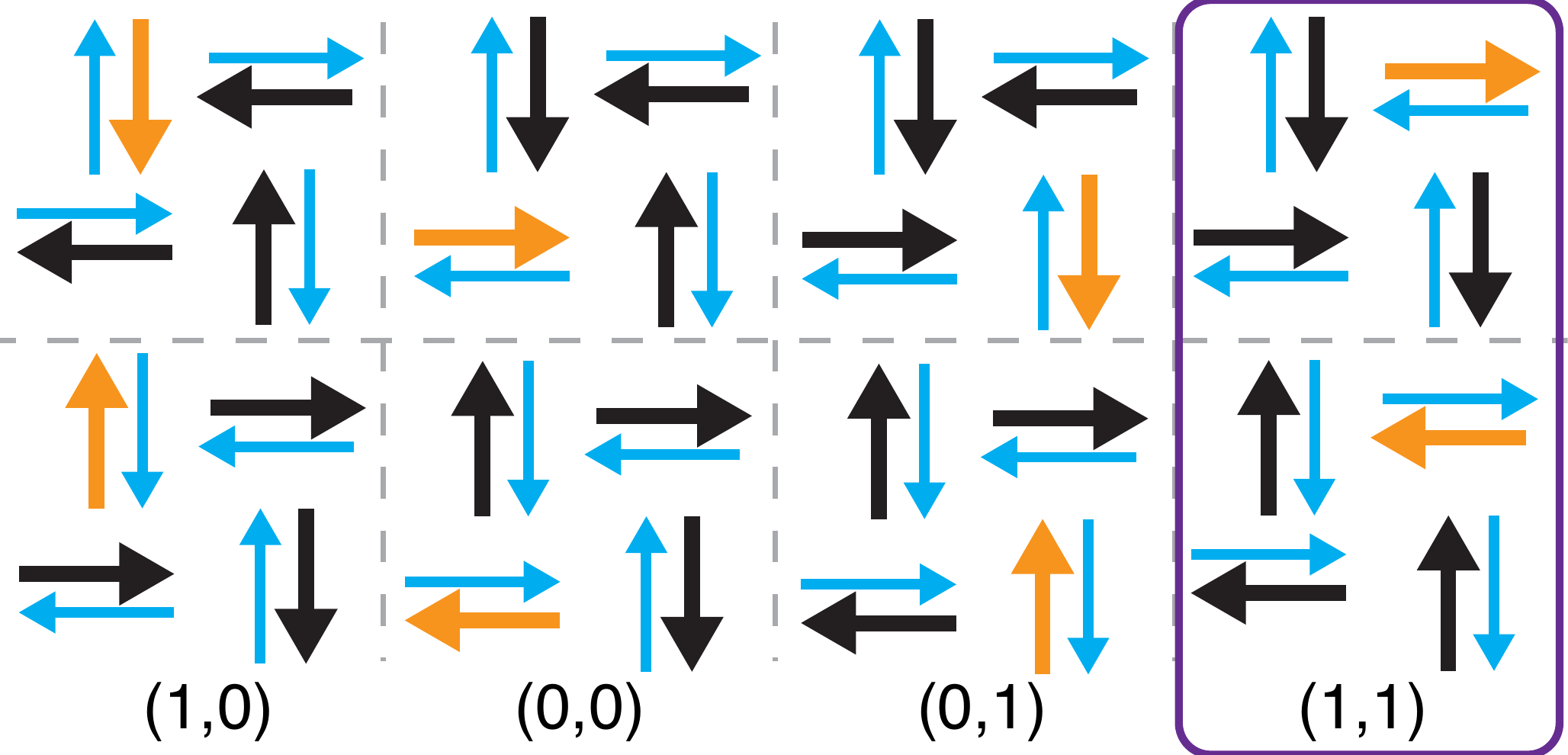}
\caption{Selection of 8 possible ordering patterns in the ground state of the short-range interaction (Eq.~(\ref{tan_local})) as parameterized in model III. The top $xy$ plane is shown by the larger arrows, and the underlying (anti-ferroelectric) layer is shown in blue. To highlight the relative changes one molecule is colored orange. The in-plane polarization vectors are indicated at the bottom. Inclusion of long-range d-d coupling makes the states marked with purple the ground state.}
\label{gs}
\end{figure}

In Figure~\ref{gs}, eight out of a total of $\Omega=1024$ possible ground states of \textit{only} the short-range interaction (Eq.~(\ref{tan_local})) as parameterized in model III are sketched. The most in-plane polar states (highlighted in purple) that alternate between (1,1) and ($-$1,$-$1) polarized layers is the expected ordering. The inclusion of NN dipole-dipole coupling splits the degeneracy of the GS and brings its multiplicity down from $1024\rightarrow32$. It couples the layers in the $z$-direction to each other and enforces their AFE ordering. Still all states of Fig.~\ref{gs} and further symmetry equivalents are possible ground states. If the cutoff radius of the d-d interaction is extended to include the next NN ($r_c=\sqrt{2}a$), the GS of Fig.~\ref{gs} splits into three energy levels. Whereby the (1,1) state is the new GS with $\Omega=8$, followed in energy by the (1,0) and (0,0) state. This means that d-d interaction in model III stabilizes the long-range ordering pattern to $\mathbf{M}=(0.5,0.5,1)$ and makes it the GS. Note that including only NN coupling would enable a checkerboard pattern of the different orderings (shown in Fig.~\ref{gs}) at low temperature.

\begin{figure}[!b]
\centering
\includegraphics[width=\columnwidth]{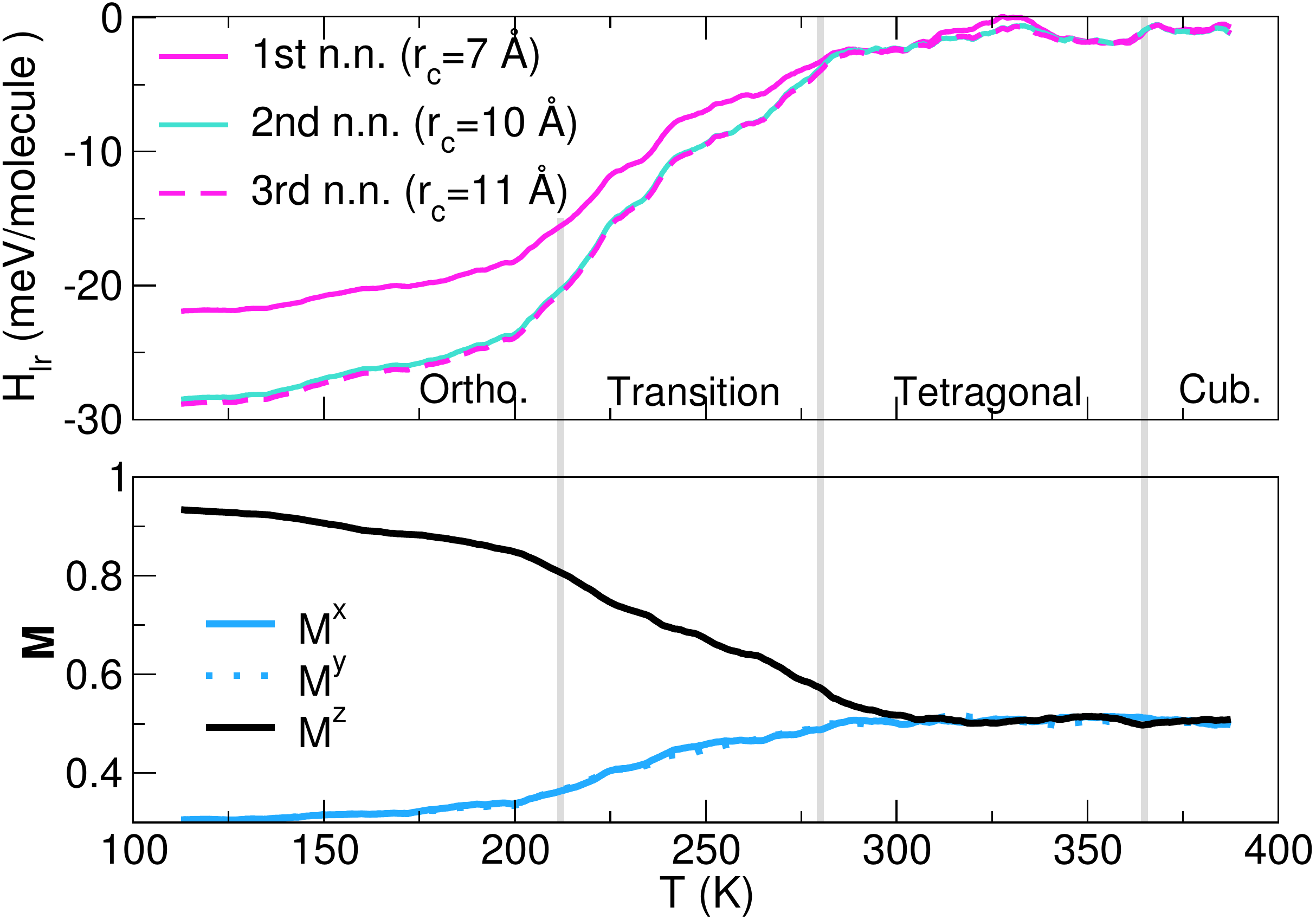}
\caption{The long-range energy ($H_{\rm lr}$, $\varepsilon_r=1$) and order parameter ($\mathbf{M}$) calculated based on the molecular C-N axes extracted from a 0.5~K/ps heating run with the MLFF. The trajectory is started from a $4\times4\times4$ supercell in the orthorhombic phase and run in the $NPT$ ensemble. Running averages with a width of 25~K are applied to smoothen the curves. The vertical gray lines sketch boundaries between the different crystal phases.  }
\label{Epdheat}
\end{figure}

We now turn the problem around and use the unscreened long-range energy ($H_{\rm lr}$, $\varepsilon_r=1$), to analyze the ordering patterns obtained with the MLFF. To calculate Eq.~(\ref{dipdip}), the center and direction of the C-N axes are used in combination with the constant value for $|\mathbf{p}|$. In Figure~\ref{Epdheat} the energy and molecular order parameter are shown for a system in which the temperature was continuously raised from 100~K to 400~K. The still fairly fast heating rate results in hysteresis spreading the orthorhombic to tetragonal transition over a $50$~K temperature range. The orthorhombic order $\mathbf{M}_{\rm ortho}$ is stabilized by d-d interactions by $\sim30$~meV/molecule, whereas this contribution drastically reduces in the tetragonal phase and vanishes in the cubic phase. This trend is the same in constant temperature MDs, where we obtain on time-average -29, -3.7, -2.6 and -1.7~meV/molecule at 150, 250, 300 and 400~K, respectively. These unscreened energies from an upper-bound to the real d-d interaction energy. This indicates that long-range d-d interactions stabilize the orthorhombic phase, but essentially play no role in the more disordered tetragonal and cubic phases.


\subsection{Accuracy of the long-range interaction}

\subsubsection{The MA$_x$Cs$_{1-x}$PbI$_3$ test system}

The pd-interaction plays an essential role in the stabilization of the long-range molecular order. However, the origin and validity of the selected parameters in the models is unclear. To test their accuracy, we have constructed MA$_x$Cs$_{1-x}$PbI$_3$ test systems and examined the dipole-dipole interactions by DFT calculations. The systems are based on $2\times2\times2$ CsPbI$_3$ cubic supercells as shown in Figure~\ref{fig:Csframe}. The central Cs atom is replaced by a MA molecule which is rotated manually. The DFT total energy ($E$) of the rotated configurations $\{0\leq\phi<\nicefrac{\pi}{2},0\leq\theta<\pi\}$ is used to map out the short-range interaction energy $E_a(\phi,\theta)$. The obtained energy surface has a corrugation energy of $\sim80$~meV with a room-diagonal orientation in the cubic PbI framework that is unfavored. Hereafter, a second Cs is replaced by a fixed MA molecule to probe the d-d coupling. As before, the central molecule is rotated, and the DFT total energies are calculated. Depending on the position of the fixed molecule, see Figs.~\ref{fig:Csframe}~(b) axes $\mathbf{n}=(\pm1,0,0)$ and (c) room diagonal $\mathbf{n}=(\pm1,\pm1,\pm1)$ neighbors, energy surfaces $E_b(\phi,\theta)$ and $E_c(\phi,\theta)$ are obtained, respectively. Assuming the short-range interaction is unaffected by the replacement of a neighboring Cs by MA, we can extract the long-range interaction energy by calculating
\begin{equation}
 \Delta{}E_{x}=E_{x}-E_a-\langle{}E_{x}-E_a\rangle{},
 \label{DFTE}
\end{equation}
where $x=\{b,c\}$ and $\langle{}.\rangle$ the average over $\{\theta,\phi\}$. The DFT long-range interaction energy obtained in test systems (b) and (c) (shown on the bottom of Fig.~\ref{fig:Csframe}) are accurately fitted by: $\Delta{}E_{x}=\alpha_x\cos{\theta}$, where $\alpha_b=-35$~meV and $\alpha_c=-10$~meV. In the pd-approximation these systems are described by $\Delta{}E^{\rm pd}_{x}=H_{\rm lr}$ using periodic-boundaries. Summing the interaction in real space and cutting off the interactions beyond $r_c=\sqrt{3}a$, the following two energies are obtained 
\begin{equation}
 \Delta{}E^{\rm pd}_{b}=-4\frac{C}{a^3\varepsilon_{r}}\cos{\theta},\quad \Delta{}E^{\rm pd}_{c}=0,
 \label{pdbc}
\end{equation}
where $C=\frac{|\mathbf{p}|^2}{4\pi \varepsilon_{0}}$ is the Coulomb constant. Setting $|\mathbf{p}|$ to 2.29~D fixes the energy scale to $C/a^3\approx13$~meV. This means that the pd model with the real space sum and $\varepsilon_r=\frac{4\times13}{\alpha_b}\approx1.5$ describes the DFT result for system~(b). However, it is unable to capture the cosine with amplitude $\alpha_c$ observed in system~(c), even when the interaction radius is increased.

\begin{figure}
 \centering
 \includegraphics[width=.9\linewidth]{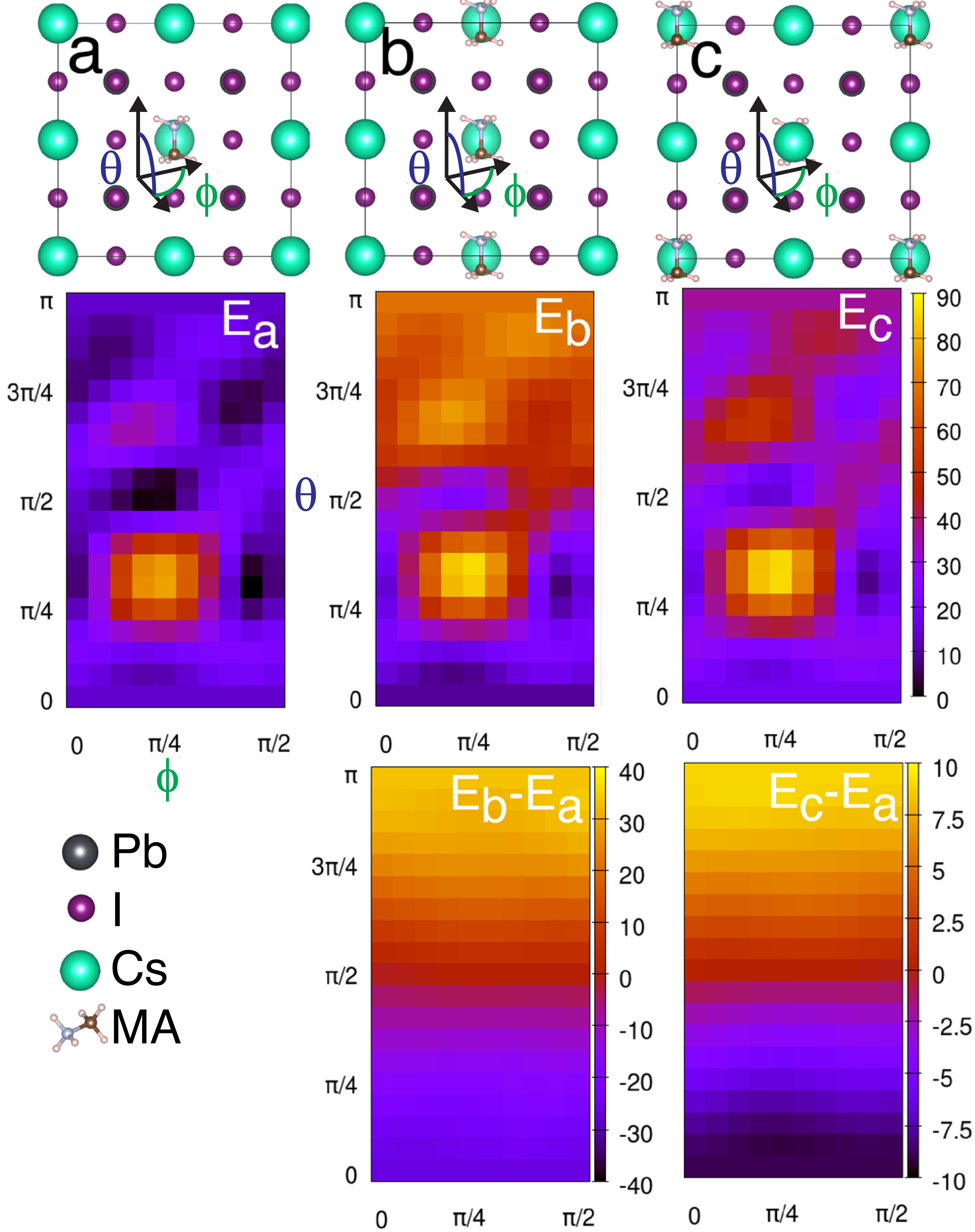}
 \caption{\textit{top:} The MA$_x$Cs$_{1-x}$PbI$_3$ test system in a $2\times2\times2$ cubic supercell. a) One MA molecule rotated by $\{\phi,\theta\}$. b) A fixed MA molecule at distance $r=a$ and direction $\hat{\mathbf{n}}=(0,0,1)$ replaces the Cs. c) Same as in b), but with $r=\sqrt{3}a$ and $\hat{\mathbf{n}}=\nicefrac{1}{\sqrt{3}}(1,1,1)$. \textit{middle:} DFT total energy surfaces of systems a, b and c. \textit{bottom:} DFT energy differences ($\Delta{}E_{b},\,\Delta{}E_{c}$). The energy scale on all colorbars is in meV.}
 \label{fig:Csframe}
\end{figure}

\begin{table}
\centering
\caption{Parameters of the model dielectric function. Fitted 
screening length parameter ($\lambda$) in $\angstrom^{-1}$ and the "ion clamped" static 
dielectric constant ($\varepsilon_{\infty}$) both adapted from Ref.~\onlinecite{Bokdam:sr16}.}
\label{tb:GW}
\begin{tabular}{lcclcc}
& $\lambda$& $\varepsilon_{\infty}$&& $\lambda$& $\varepsilon_{\infty}$\\ \hline
MASnI$_3$&1.05&9.18&MAPbBr$_3$&1.13&5.15\\
FASnI$_3$&1.05&8.06&FASnBr$_3$&1.13&5.32\\
FAPbI$_3$&1.05&7.10&FAPbCl$_3$&1.17&4.27\\
MAPbI$_3$&1.05&6.83&MAPbCl$_3$&1.17&4.22\\
MASnBr$_3$&1.13&5.89&FASnCl$_3$&1.17&4.07\\
FAPbBr$_3$&1.13&5.25&MASnCl$_3$&1.17&4.05\\

\end{tabular}
\end{table}

\begin{figure}[!t]
 \centering
 \includegraphics[width=\columnwidth]{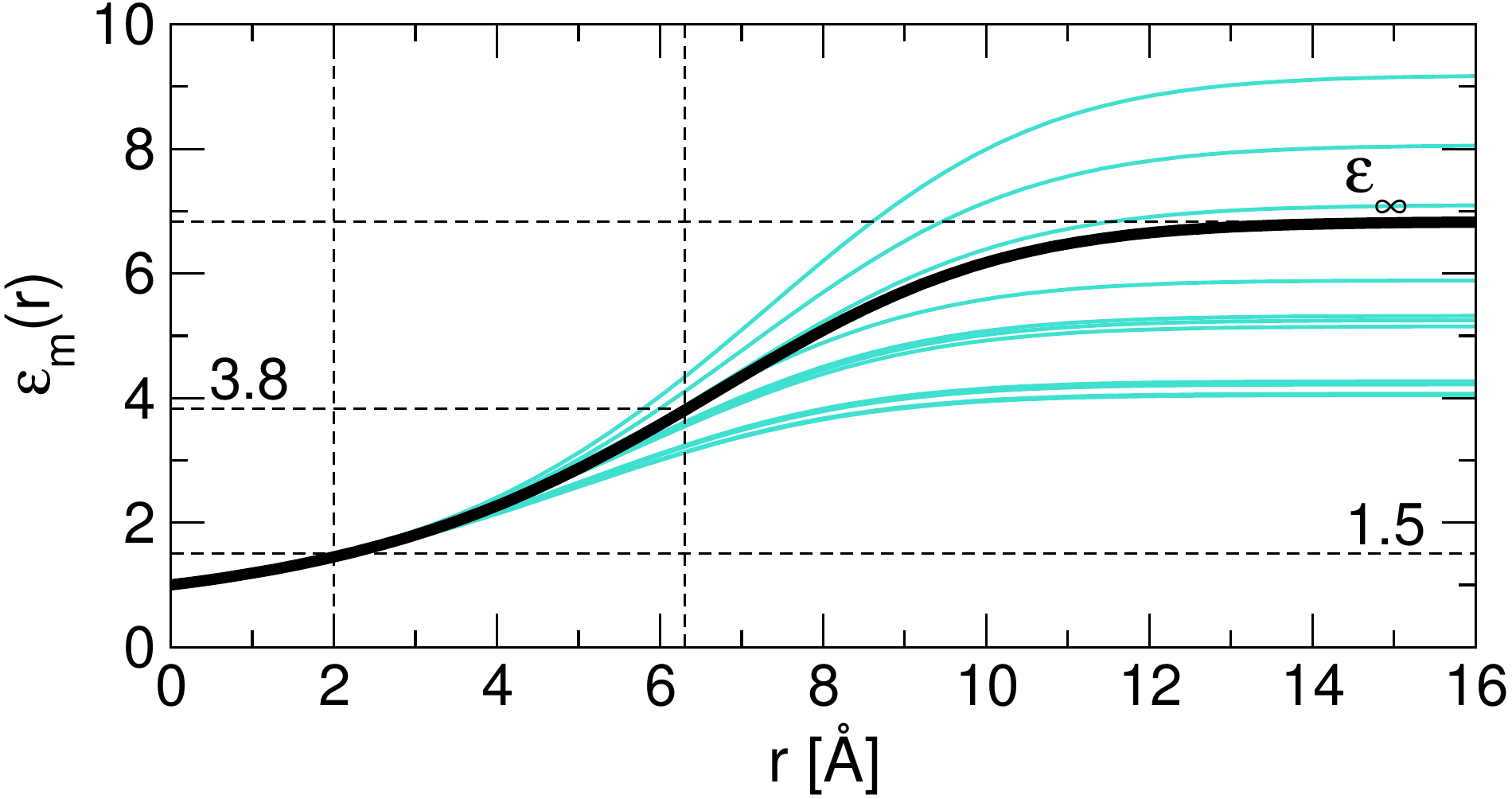}
 \caption{Distance dependent dielectric function $\varepsilon{}_{\rm m}(r)$ for MAPbI$_3$ in the cubic phase based on the parameters of Table~\ref{tb:GW}. The range of screening strengths in the wider class of hybrid perovskites is plotted by the turquoise lines.}
 \label{fig:eps}
\end{figure}

\subsubsection{Changes to the long-range interaction}
\label{changes}

In order to understand the missing coupling in system (c), we further analyze the d-d interaction using three different electrostatic models.

First, considering the long-range nature of Eq.~(\ref{dipdip}), the conditional convergence of Eq.~(\ref{long_range}) under the periodic boundaries that are imposed\cite{Markov:prb95}, $H_{\rm lr}$ should be calculated by an Ewald summation. Here we apply the formulation of the Ewald sum as presented in Ref.~\onlinecite{FRENKELBOOK}. Note that the ($\mathbf{k}=0$) term in the Ewald sum is omitted, implying that "conducting boundaries" are assumed\cite{Shougang:prb11}. 

Second, the microscopic electronic screening in this semiconductor can be more accurately described. Here, we apply the following distance dependent dielectric function
\begin{equation} 
  \varepsilon_{\rm m}^{-1}(r)=1-(1-\epsilon^{-1}_{\infty})\text{erf}\left ({\color{black}\frac{\lambda}{2\pi}} r\right ),
  \label{modelepsr}
\end{equation}
where $r$ is the distance, and ${\color{black}\frac{\lambda}{2\pi}}$ is the real space screening length. Equation~(\ref{modelepsr}}) is the real space version of the local model dielectric function fitted to the Many-Body Perturbation Theory calculations in the $GW_0$ approximation as described in the supplementary materials of Ref.~\onlinecite{Bokdam:sr16} {\color{black} It was used there to screen electron-hole interactions in MAPbI$_3$.} The Fourier transform of $\varepsilon^{-1}(|\mathbf{k}+\mathbf{G}|)\rightarrow{}\varepsilon^{-1}(r)$ is described in Appendix~\ref{AppEps}. Compared to one fixed effective dielectric constant ($\varepsilon_{r}$) this is a more realistic screening description with proper limits, $\varepsilon(0)=1$ and $\varepsilon(\infty)=\varepsilon_\infty$. {\color{black} This screening function is closely related to the hybrid screened exchange functional, where $\lambda$ is the range-separation parameter\cite{Liu:arxiv19}.} Note it only includes contributions from electronic screening, the ionic contributions are not included. We tabulate the parameters for a set of commonly studied hybrid perovskites in Table~\ref{tb:GW} and plot the inverse of $\varepsilon^{-1}_{\rm m}(r)$ in Figure~\ref{fig:eps}. We recall the $\varepsilon_{r}$ values for the three models $\varepsilon^{\rm I}_{r}=1$, $\varepsilon^{\rm II}_{r}=0.73$, $\varepsilon^{\rm III}_{r}=1.5$. Since the dominant contribution to the total dipole-dipole interaction energy comes from NN coupling, which occurs at distances of one lattice spacing ($\sim$6.3~\angstrom), setting $\varepsilon_{r}<\varepsilon_{\infty}$ as an effective screening parameter seems reasonable. {\color{black} In comparison to models I-III, the distance dependent dielectric function screens more, with a NN screening of $\varepsilon_{\rm m}(a)=3.8$. }

Third, the use of the pd-approximation can be circumvented by replacing the atoms with point charges, thereby creating interacting monopoles (mm). The system is sketched in Fig.~\ref{structures} and is in total charge neutral. The same dipole moment as before is used to fix the charges: $q_{\rm C}=2.29\times0.20819434\,{\rm e\AA{}}/1.51\,{\rm \AA{}}=0.316{\rm e}$, $q_{\rm N}=-q_{\rm C}$, $q_{\rm Pb}=-2q_{\rm I}$, $q_{\rm H}=0$ and $q_{\rm MA}=q_{\rm Cs}=-q_{\rm I}$. A compensation charge $q_{\rm MA}$ is placed on the middle of the C-N bond. The Ewald sum is used to calculate the electrostatic energy of the monopole arrangements. Since the PbI framework and the center of mass of the molecule are fixed in this analysis, the charges $q_{\rm Pb}$, $q_{\rm I}$ and $q_{\rm MA}$ only create a potential offset.

\begin{figure}[!t]
 \centering
 \includegraphics[width=\linewidth]{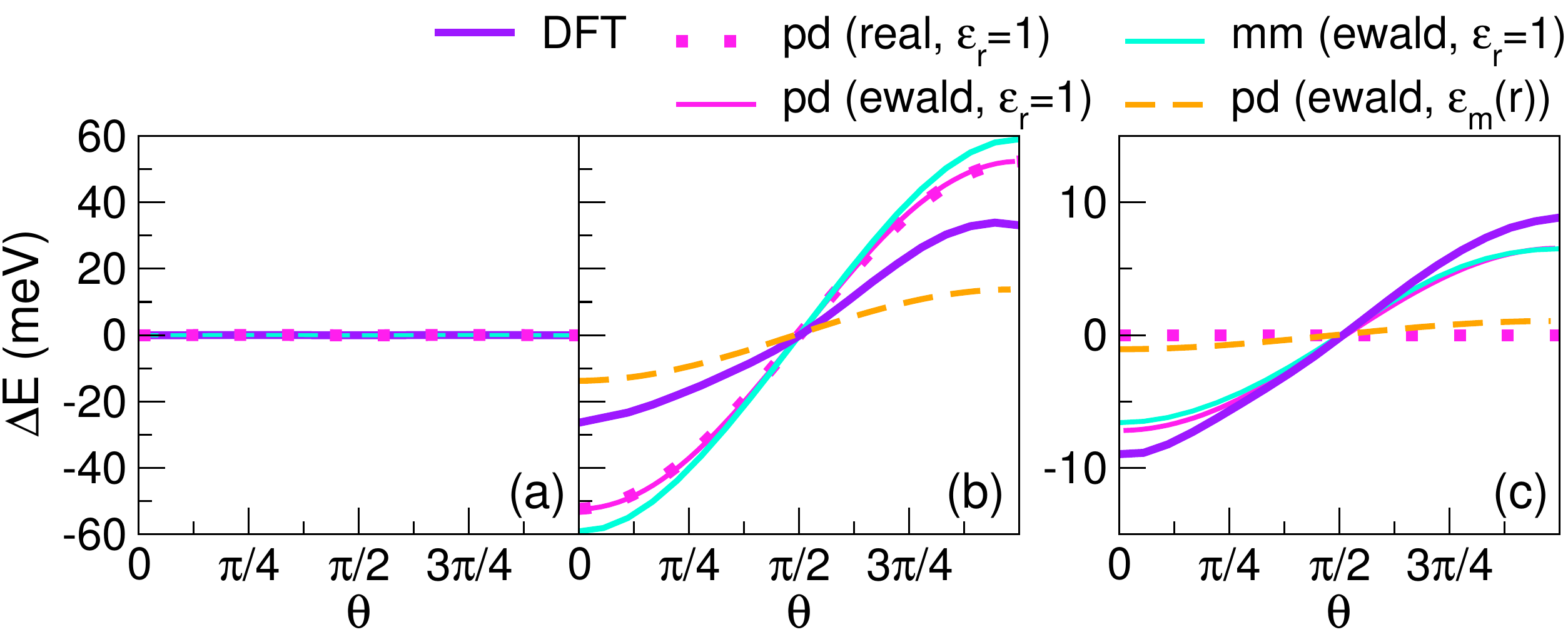}
 \caption{Long-range interaction energy calculated in the point-dipole (pd) and monopole-monopole (mm) approximation using the Ewald method or the real space sum. Both constant screening ($\varepsilon_r$) and the distance dependent screening function ($\varepsilon_{\rm m}(r)$) are applied. Results are compared to the DFT data ($\Delta{}E_{x}$) for the test systems (a), (b) and (c) in the respective figures.}
 \label{fig:barr}
\end{figure}

In Figure~\ref{fig:barr}, we have applied the three changes to the test systems (a)-(c). The most clear improvement is that by applying the Ewald sum the cosine term is recovered for system (c). This term appears from the reciprocal space sum in the Ewald method. Note that the coupling to the second NN $\mathbf{n}=(\pm1,\pm1,0)$ are correctly accounted for in the real-space sum. 
The mm-model (solid cyan line) shows almost the same result as the pd-model with the Ewald sum. This means that the distance between C and N is sufficiently small compared to the NN distance $a$. The cosine behavior of both systems (b) and (c) are captured by the pd-model with Ewald sum and $\varepsilon_r=1$ , but their amplitudes are not. In system (b) the coupling is overestimated, while it underestimates the coupling in system (c). {\color{black} This is not mitigated by the distance dependent dielectric function $\varepsilon_{\rm m}(r)$, which "over-screens" the coupling in both systems. The amplitude of the DFT energy is approximately 2$\times$ and 8$\times$ larger than the $\varepsilon_{\rm m}(r)$ screened d-d interaction energy of systems (b) and (c), respectively.} Overall, none of the applied electrostatic schemes can capture the strength of the interactions calculated by DFT in both systems simultaneously.

\section{Discussion}
\label{sec:dis}
With the Cs/MA-testsystem we removed what in most model studies of MAPbI$_3$ would be considered as the physical/chemical reasons for the short-range interactions. \textit{Stress and strain} on the framework are removed because the framework is kept rigid under the MA rotation and the reference energy of system (a) is subtracted. We assume that the \textit{hydrogen bonds} of the molecules with the framework are only marginally affected by the rotation of a neighboring molecule and should therefore be taken away in this scheme. The pd-approximation using the Ewald method qualitatively captures the interaction between the molecules. The inability to capture the correct amplitude of the interaction in both systems (b) and (c) simultaneously can be explained by microscopic polarization effects. $H_{\text{lr}}$ assumes a constant dipole moment ($|\mathbf{p}|$), however in reality it is not constant but the result of an effective (depolarisation) field imposed by the surrounding ions and dipoles. These effects could possibly be captured by an anisotropic dielectric function. 

By including a distance dependent electronic screening function with a relatively large value for $\varepsilon_{\infty}=6.31$ we have increased the total screening of the d-d coupling compared to the previous models, especially for the second NN and beyond. The ionic screening from the PbI lattice has been omitted. In a previous study we have shown that the dielectric functions at 0~K calculated with density functional perturbation theory and at 300~K calculated with FPMD are very similar\cite{Bokdam:sr16}. The dominant (dipole-active) oscillators related to Pb-I bonds lie between 2-10 meV (2-0.4 ps), ie. slightly faster than the MA reorientation times (several ps). This results in a sizable increase of the static dielectric constant ($\varepsilon_0=30$) compared to the ‘ion-clamped’ high frequency dielectric constant ($\varepsilon_{\infty}$). It can therefore not be excluded that the d-d interactions are (partially) screened by the PbI lattice. This implies that the here computed d-d interaction energy is an upper bound.

\begin{figure}[!t]
 \centering
 \includegraphics[width=.9\linewidth]{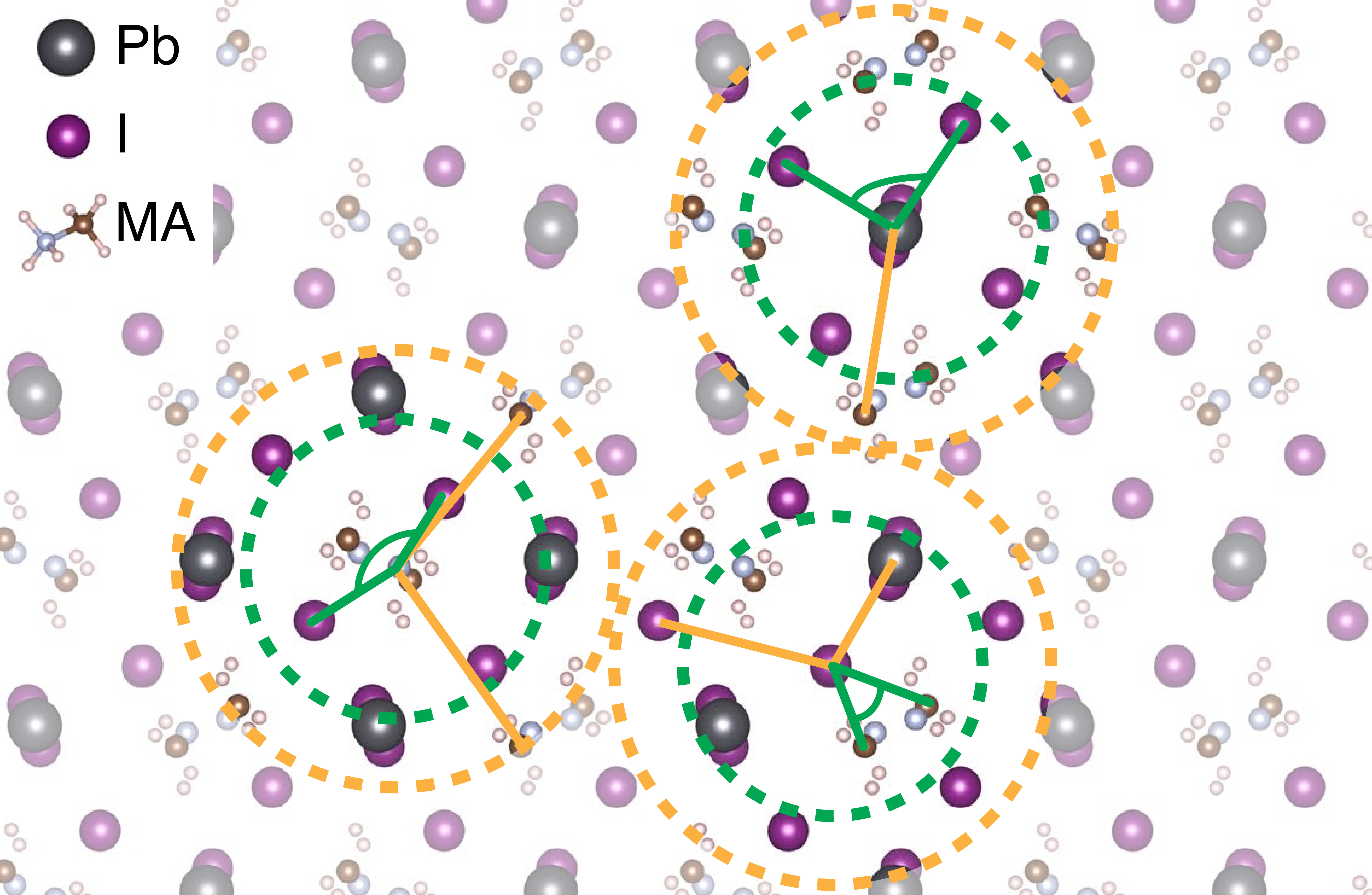}
 \caption{Local chemical environment around a lead, nitrogen and iodine atom as recognized in the machine learning force field. Two-body radial and three-body angular descriptors with a cut-off of 6~\angstrom{} and 4~\angstrom{} are sketched in orange and green, respectively.}
 \label{mlffmodel}
\end{figure}

The d-d coupling between the molecules is only an effective description of the electrostatics related to the ionic nature of the perovskite lattice. Without the MA cation, cubic PbI$_3$ is metallic in a DFT calculation. Introducing a MA or Cs turns it into a semiconductor. Each molecule donates one electron to the framework, which exactly fills up the (Pb and I states dominated) valence band. To lower the electrostatic energy, each MA$^+$ moves (with its nitrogen side) to a different PbI$_3^-$. The same mechanism also occurs in CsPbI$_3$, even though the Cs cation does not have an intrinsic dipole moment. As shown in the low temperature orthorhombic phases of Figure~\ref{structures}, similar off-center displacements for the Cs cations as for the center of masses of the MA molecules occur. And, in both perovskites the PbI octahedra show a clock-wise, anti-clockwise ordering. This results in stretched and squeezed cavities in which the molecules align, because of space filling, along the extended axes. The framework imposes the long-range order of the molecules, whereas the displaced centers of masses of the cations are the result of the ionic lattice.

To capture such interactions, we have applied a machine-learning force field approach which learns a force field \textit{on-the-fly} during FPMD\cite{Jinnouchi:prl19}. The constraining effect of the PbI framework is automatically included, one of the difficult terms to model accurately. This force field is based on a purely short-range description as shown in Figure~\ref{mlffmodel}. Two-body radial and three-body angular descriptors with a cut-off of 6~\angstrom{} and 4~\angstrom{}, respectively, are used to represent the local chemical environment around each atom. The total energy is then defined as the sum of atomic energies assigned to each atom\cite{Bartok:prl10} by means of a similarity kernel. This method shows ordering patterns, lattice constants and phase transition temperatures in very good agreement with experiments. The chemical environment around a Pb atom (shown in Fig.~\ref{mlffmodel} top right) connects both NN as well as next NN molecules. Thereby it is able to impose the AFE ordering pattern of the molecules in the orthorhombic phase. It effectively has the same reach as the minimal cutoff radius ($r_c=\sqrt{2}a$) required to split the degenerate GS of the short-range part of the model Hamiltonians. This implies that long-range d-d coupling is not necessary to capture the molecular ordering.

\section{Conclusions and Outlook}
\label{sec:con}
The ordering pattern of molecules in the hybrid perovskite MAPbI$_3$ at finite temperatures as described by three different model Hamiltonians (I-III) and an accurate MD simulation have been studied. A comparison based on the molecular order parameter ($\mathbf{M}$) has shown that model III best captures the features observed in MD. Using a discrete set of molecular orientations, model III qualitatively captures the ordered (orthorhombic) to disordered (cubic) transition upon heating, involving two second order transitions. The order-disorder transition in model I is a continuous process, while in model II the system undergoes one abrupt first-order transition. The ground state ordering pattern of model III closely resembles that of the experimental orthorhombic structure, whereas both models I and II show completely different ground state order. 

In attempt to improve upon the models, we examined the accuracy of the dipole-dipole interactions using a more advanced electrostatic description with a distance dependent dielectric function and the Ewald summation technique. Although these methods enable a better quantitative description of the electrostatic interaction, the improvement was found to be not substantial. Static MA$_x$Cs$_{1-x}$PbI$_3$ test systems were constructed to probe the d-d interaction energies in self-consistent DFT calculations. The point-dipole model is able to qualitatively describe the DFT interaction energies. However, its strength can not be consistently captured with one set of screening parameters.

We have shown that the pd-approximation ($H_{\text{lr}}$) effectively splits the degenerate ground state imposed by the short-range part ($H_{\text{sr}}$) of the  Hamiltonian ($H=H_{\text{lr}}+H_{\text{sr}}$) of model III. A short cutoff radius ($r_c=\sqrt{2}a$) for the calculation of $H_{\text{lr}}$ including the first and second NN dipole already suffices to capture the long-range AFE order of the molecules observed in the low-temperature orthorhombic phase. The d-d interaction energy calculated for ordering patterns from the MLFF shows a stabilizing effect in the orthorhombic phase, however it is practically absent in the tetragonal and cubic phases. The essence of the long-range order in the low-temperature orthorhombic phase lies in the propagation of this order by short-range interactions. The MLFF effectively captures these interactions between atoms that are no more than 6~\angstrom{} apart. This analysis explains the large similarity of the MAPbI$_3$ and CsPbI$_3$ crystal lattices at low temperature, even though Cs does not have a permanent dipole-moment. 

In future work, it is possible to improve the model Hamiltonians further by adopting more elaborate descriptions of $H_{\text{sr}}$ as well as inclusion of more refined molecular degrees of freedom. However, its development and optimization \textit{by-hand} are not tractable. In contrast, the MLFF method systematically constructs an accurate interaction model \textit{on-the-fly} during FPMD. It enables the automatic inclusion of the constraining effect of the PbI framework as well as the effective short-range electrostatics with high accuracy, thereby realizing realistic ordering patterns of the molecules. The method can be extended to more complex systems with defects, surfaces and mixed halides or molecules.


\acknowledgements{We appreciate discussions with Merzuk Kaltak and Peitao Liu in finalizing the manuscript. M.B. thanks Viktor Fournarakis and Lorenzo Papa for helpful discussions. Funding by the Austrian Science Fund (FWF): P 30316-N27 is gratefully acknowledged. Computations were performed on the Vienna Scientific Cluster VSC3. R.J. appreciates the financial support from Toyota Central R\&{}D Labs., Inc.}

\appendix

\section{Real space dielectric function}
\label{AppEps}
The Coulomb potential in reciprocal space screened by the model dielectric function $\varepsilon_{\rm m}$ as presented in Ref.~\onlinecite{Bokdam:sr16} is given by
\begin{equation}
 V(\mathbf{k})=\left(\frac{4\pi}{\mathbf{k}^{2}}\right)\varepsilon^{-1}_{\rm m}(|\mathbf{k}|)
    =\frac{4\pi}{\mathbf{k}^{2}} \left(1-(1-\varepsilon^{-1}_{\infty})e^{-\frac{|\mathbf{k}|^{2}}{{\color{black}4}\lambda^{2}}}\right ).
 \label{Pot1}
\end{equation}

The screened Coulomb potential in real space is obtained by the Fourier transform of Eq.~(\ref{Pot1}),
\begin{equation}
 V(\mathbf{r})={\color{black}\frac{1}{(2\pi)^3}}\int_{-\infty}^{\infty}\left [\frac{4\pi}{\mathbf{k}^{2}} \left(1-(1-\varepsilon^{-1}_{\infty})
    e^{-\frac{|\mathbf{k}|^{2}}{{\color{black}4}\lambda^{2}}}\right ) \right ]e^{i\mathbf{kr}}d\mathbf{k}.
 \label{Pot2}
\end{equation}
Switching to spherical coordinates and evaluating the integrals with respect to the spherical angles $\theta$ and $\varphi$ gives,
\begin{equation}
 V(r)={\color{black}\frac{1}{(2\pi)^3}}\frac{16\pi^{2}}{r}\int_{0}^{\infty}\left [ \frac{1-(1-\varepsilon^{-1}_{\infty})
                 e^{-\frac{k^{2}}{{\color{black}4}\lambda^{2}}}}{k} \right ] \sin(kr) dk.
 \label{Pot3}
\end{equation}
Hereby, the problem reduces to solving the integrals
\begin{equation}
A(r)=\int_{0}^{\infty}\frac{\sin(kr)}{k}dk,
\label{Pot4}
\end{equation}
and
\begin{equation}
B(r)=\int_{0}^{\infty}
\frac{sin(kr)e^{-\frac{k^{2}}{{\color{black}4}\lambda^{2}}}}{k}dk.
\label{Pot5}
\end{equation}

To solve $A(r)$, a damping function $e^{-\beta k}$ is multiplied to the integrand. Note that in the limit of $\beta \rightarrow 0$ the original function is recovered. The 
derivative of this integral with respect to $\beta$ is
\begin{equation}
\frac{d I(\beta,r)}{d \beta}=-\int_{0}^{\infty}\sin(kr)e^{-\beta k}dk=-\frac{r}{\beta^{2}+r^{2}}.
\label{Pot6}
\end{equation}
The indefinite integral with respect to $\beta$ results in the inverse tangent
function $I(\beta,r)=-\arctan\left (\frac{\beta}{r} \right)+C$. By considering the limit $\beta \rightarrow \infty$,
\begin{equation}
 I(\infty,r)=-\frac{\pi}{2}+C=\int_{0}^{\infty}\frac{\sin(kr)}{k}e^{-\infty k}dk=0,
 \label{Pot7}
\end{equation}
one can determine the integration constant $C=\pi/2$. Thereby Eq.~(\ref{Pot4}) reduces to  
\begin{equation}
A(r)=\frac{\pi}{2}.
\label{Pot8}
\end{equation}

To solve $B(r)$, one introduces $D=\frac{1}{{\color{black}4}\lambda^{2}}$ and takes the derivative with respect to r,
\begin{equation}
\frac{dB(r)}{dr}=\int_{0}^{\infty}e^{-Dk^{2}}cos(kr)dk=\frac{1}{2}\int_{-\infty}^{\infty}e^{-Dk^{2}}e^{ikr}dk.
 \label{Pot9}
\end{equation}
The integration over k simplifies to a 1-dimensional Fourier transform of a Gaussian function,
\begin{equation}
 \frac{dB(r)}{dr}=\frac{1}{2}{\color{black}\sqrt{\frac{\pi}{D}}}e^{-\frac{r^{2}}{4D}}.
 \label{Pot10}
\end{equation}
Integrating with respect to r results in
\begin{equation}
 B(r)=\frac{\pi}{2}\text{erf}\left (r\lambda \right ).
 \label{Pot11}
\end{equation}
Combining the Eqs.~(\ref{Pot8}) and (\ref{Pot11}) with Eq.~(\ref{Pot3}) gives the real space screened Coulomb potential
\begin{equation}
 V(r)={\color{black}\frac{1}{(2\pi)^3}}\frac{16\pi^{2}}{r}\frac{\pi}{2} \varepsilon_{\rm m}^{-1}(r)={\color{black}\left(\frac{1}{r}\right)\varepsilon_{\rm m}^{-1}(r)},
\end{equation}
where 
\begin{equation}
 \varepsilon_{\rm m}^{-1}(r)=1-(1-\epsilon^{-1}_{\infty})\text{erf}\left ( r\lambda \right ).
 \label{Pot12}
\end{equation}
 is the inverse dielectric function in real space. {\color{black} Here $\lambda$ is expressed in (angular) wavenumbers.}

\end{document}